\documentclass[twocolumn,aps,pra,longbibliography,sort&compress,showpacs,amsfonts,amsmath,amssymb,superscriptaddress,footinbib,floatfix]{revtex4-1}

\usepackage[latin9]{inputenc}
\usepackage{amsmath}
\usepackage{amssymb}
\usepackage{graphicx}
\usepackage{color}

\makeatletter

\usepackage{ulem}
\newcommand{\kbar}{\mathchar'26\mkern-9mu k}

\makeatother

\begin{document}

\title{Bogoliubov excitations in the quasiperiodic kicked rotor: stability of a kicked condensate and the quasi-insulator-metal transition}

\author{Beno\^it Vermersch}

\affiliation{Universit\'e de Lille, CNRS, UMR 8523 \textendash{} PhLAM \textendash{} Laboratoire de Physique des Lasers Atomes et Mol\'ecules, F-59000 Lille, France}

\author{Dominique Delande}

\affiliation{Laboratoire Kastler Brossel, Sorbonne Universit\'e, CNRS, ENS-PSL University, Coll\`ege de France; 4 Place Jussieu, 75005 Paris, France}

\author{Jean Claude Garreau}

\affiliation{Universit\'e de Lille, CNRS, UMR 8523 \textendash{} PhLAM \textendash{} Laboratoire de Physique des Lasers Atomes et Mol\'ecules, F-59000 Lille, France}
 
\begin{abstract}
We study the dynamics of a Bose-Einstein condensate in the quasiperiodic
kicked rotor described by a Gross-Pitaevskii equation with periodic boundary conditions.
As the interactions are increased, Bogoliubov excitations appear and
deplete the condensate; we characterize this instability by considering
the population of the first Bogoliubov mode, and show that it does
not prevent, for small enough interaction strengths, the observation
of the transition. However, the predicted subdiffusive behavior is
not observed in the stable region. For higher interaction strengths,
the condensate may be strongly depleted before this dynamical regimes set in.
\end{abstract}


\maketitle

\section{Introduction}

Ultracold atoms are clean, controllable, and flexible systems whose
dynamics can be modeled from first principles. Interacting ultracold
bosons are often well described by a mean-field approximation leading
to the Gross-Pitaevskii equation (GPE)~\cite{Dalfovo:BECRevTh:RMP99,Cohen-TannoudjiDGO:AdvancesInAtomicPhysics::11},
which is useful in many situations of experimental interest: Superfluidity
and vortex formation~\cite{Madison:VortexFormationBEC:PRL00}, chaotic
behavior~\cite{Thommen:ChaosBEC:PRL03,Fallani:InstabilityBEC:PRL04,
Gattobigio:ChaoticPotentialsMatterWave:PRL11,Vermersch:EmergenceOfNonlinearity:PRA15},
soliton propagation~\cite{Khaykovich:MatterWaveBrightSoliton:S13},
etc. Ultracold-atom systems are thus increasingly used to realize
simple models that are inaccessible experimentally in other areas
of physics~\cite{Bloch:QuantumSimulationsUltracoldGases:NP14}.

Ultracold gases in a disordered optical potential have been used as
an emulator for the Anderson model~\cite{Anderson:LocAnderson:PR58},
allowing the direct observation of the Anderson localization~\cite{Kondov:ThreeDimensionalAnderson:S11,Jendrzejewski:AndersonLoc3D:NP12,
Semeghini:MobilityEdgeAnderson:NP15}.
The quantum kicked rotor (QKR), obtained by placing cold atoms in
a pulsed standing wave, is also a (less obvious) quantum simulator
for the Anderson physics~\cite{Fishman:LocIncommPot:PRL82,Casati:LocDynFirst:LNP79}:
It displays \emph{dynamical localization}, a suppression of chaotic
diffusion in the momentum space, recognized to be equivalent to the
Anderson localization~\cite{Fishman:LocIncommPot:PRL82}. Recent studies suggest that interactions (treated
in the frame of the GPE) lead to a progressive destruction of the
dynamical localization, which is replaced by a subdiffusive regime~\cite{Shepelyansky:KRNonlinear:PRL93,Gligoric:InteractionsDynLocQKR:EPL11,Rebuzzini:EffectsOfAtomicInteractionsQKR:PRA07,Cherroret:AndersonNonlinearInteractions:PRL14}
in analogy with what is numerically observed for the 1D Anderson model with bosonic mean-field interactions 
itself~\cite{Pikovsky:DestructionAndersonLocNonlin:PRL08,Laptyeva:DisorderNonlineChaos:EPL10,Vermersch:AndersonInt:PRE12}.

Applying standing-wave pulses (kicks) to a Bose-Einstein condensate
(BEC) may lead to a dynamical instability which transfers atoms from
the condensed to the non-condensed fraction, a phenomenon which is
not described by the GPE. The most common correction to GPE in this
context is the Bogoliubov-de Gennes (BdG) approach~\cite{Bogoliubov:TheTheoryOfSupefluydity:JPURSS47,deGennes:SuperconductivityOfMetals:66}.
The BdG theory considers ``excitations'' \textendash{} described
as independent bosonic quasiparticles \textendash{} of the Bose gas,
and thus indicates how (and how much) it differs from a perfectly
condensed gas. It has been applied both to the description of the
dynamical instability in the periodic kicked rotor~\cite{Zhang:TransitionToInstabilityKickedBEC:PRL04,Billam:CoherenceAndInstabilityInADrivenBEC:NJP12,Reslen:DynamicalInstabilityInKickedBEC:PRA08}
and to the study of a one-dimensional weakly interacting BEC~\cite{Lugan:AndersonLocalizationBogolyubov:PRL07,Gaul:BogoliubovExcitationsOfDisorderedBEC:PRA11}
in a disordered potential. In the latter case it was found that the
quasiparticles may also display Anderson localization. Interestingly,
a modified version of the QKR, the \emph{quasiperiodic} kicked rotor
(QPKR), emulates, in the absence of interactions, the dynamics of
a 3D Anderson-like model, and displays the Anderson metal-insulator
transition~\cite{Casati:IncommFreqsQKR:PRL89,Shepelyansky:3DRandomPot:JPI96}.
With this system a rather complete theoretical and experimental study
of this transition has been performed ~\cite{Chabe:Anderson:PRL08,Lemarie:CriticalStateAndersonTransition:PRL10,Lopez:ExperimentalTestOfUniversality:PRL12,Lopez:PhaseDiagramAndersonQKR:NJP13}.
In the present work we use the Bogoliubov approach to the QPKR to
the study the stability of the condensate and to assess the possibility
of the observation of the ``quasi-insulator-metal transition'' that
replaces the Anderson localization in presence of interactions, the
localized state being replaced by a sub-diffusive one~\cite{Cherroret:AndersonNonlinearInteractions:PRL14}.
We show that for weak enough interactions, the condensate remains
stable for experimentally relevant times, and that the Bogoliubov
quasiparticles also display a phase transition. This shows that the
transition can be approached from the low interaction limit by increasing
the nonlinearity and the observation times, opening a way to its experimental
observation with the quasiperiodic quantum kicked rotor in presence
of interactions.

\section{Dynamics of Bogoliubov excitations}

\begin{figure*}
\begin{centering}
\includegraphics[height=6cm]{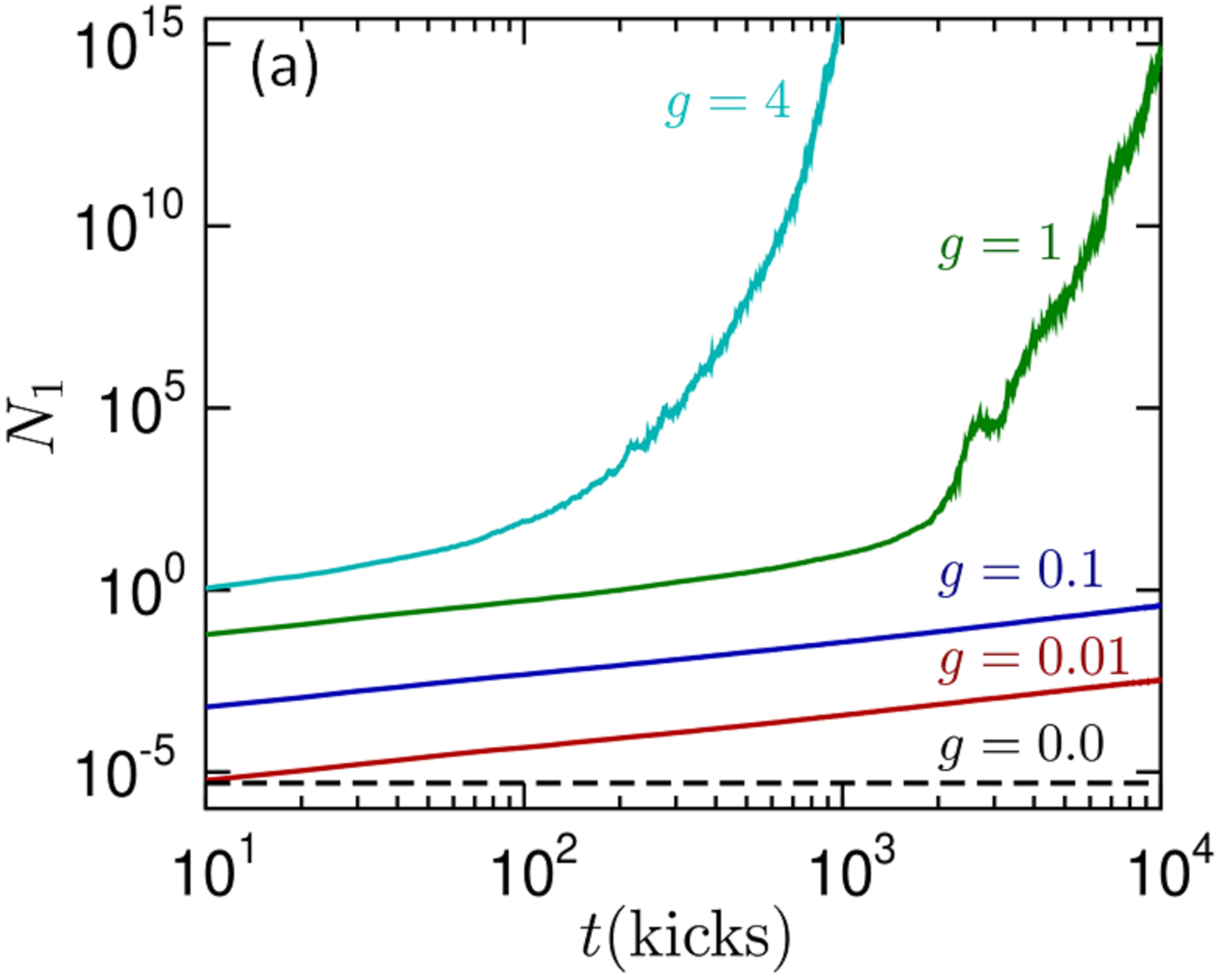}$\quad$\includegraphics[height=6cm]{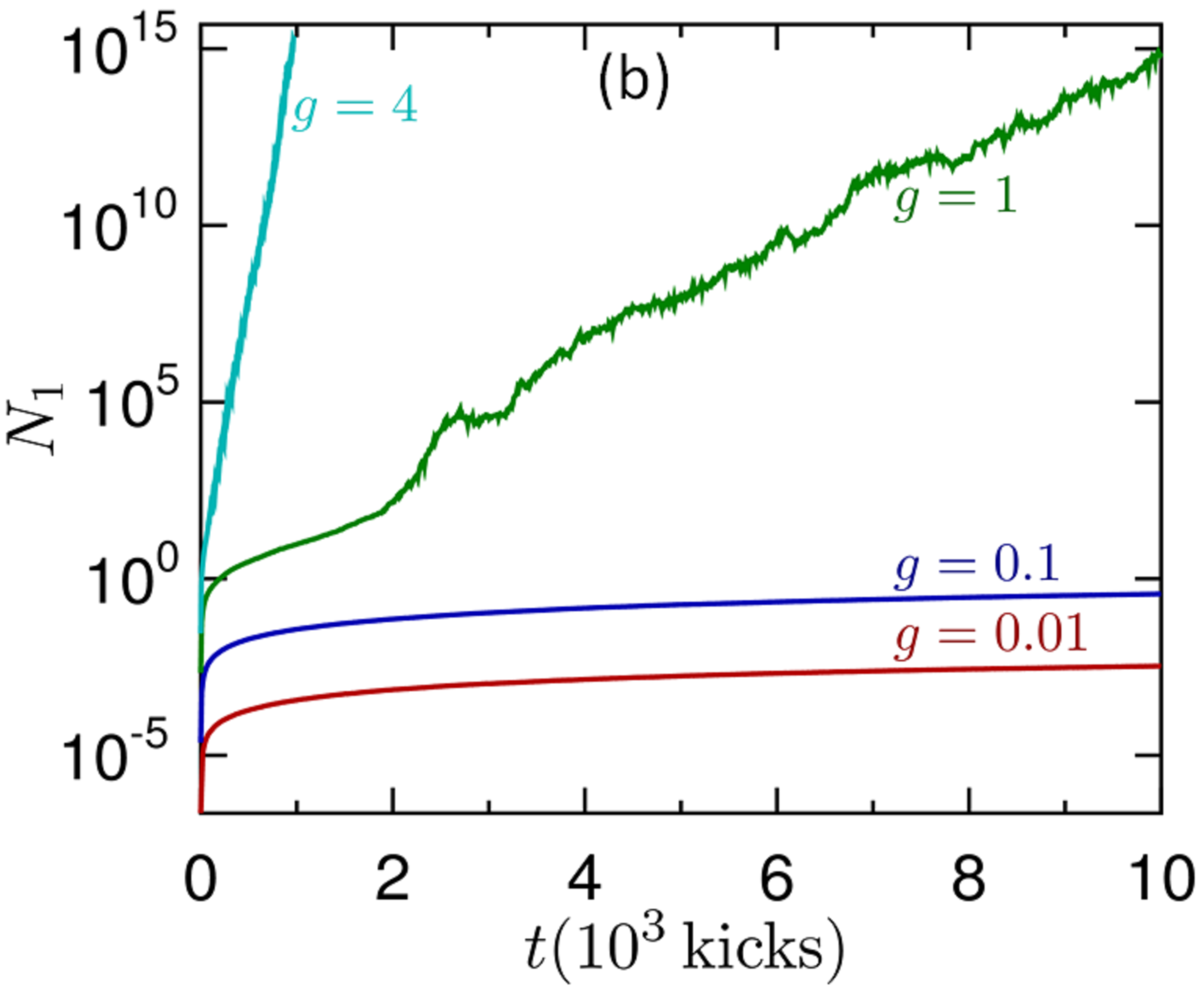}\\
\includegraphics[height=6cm]{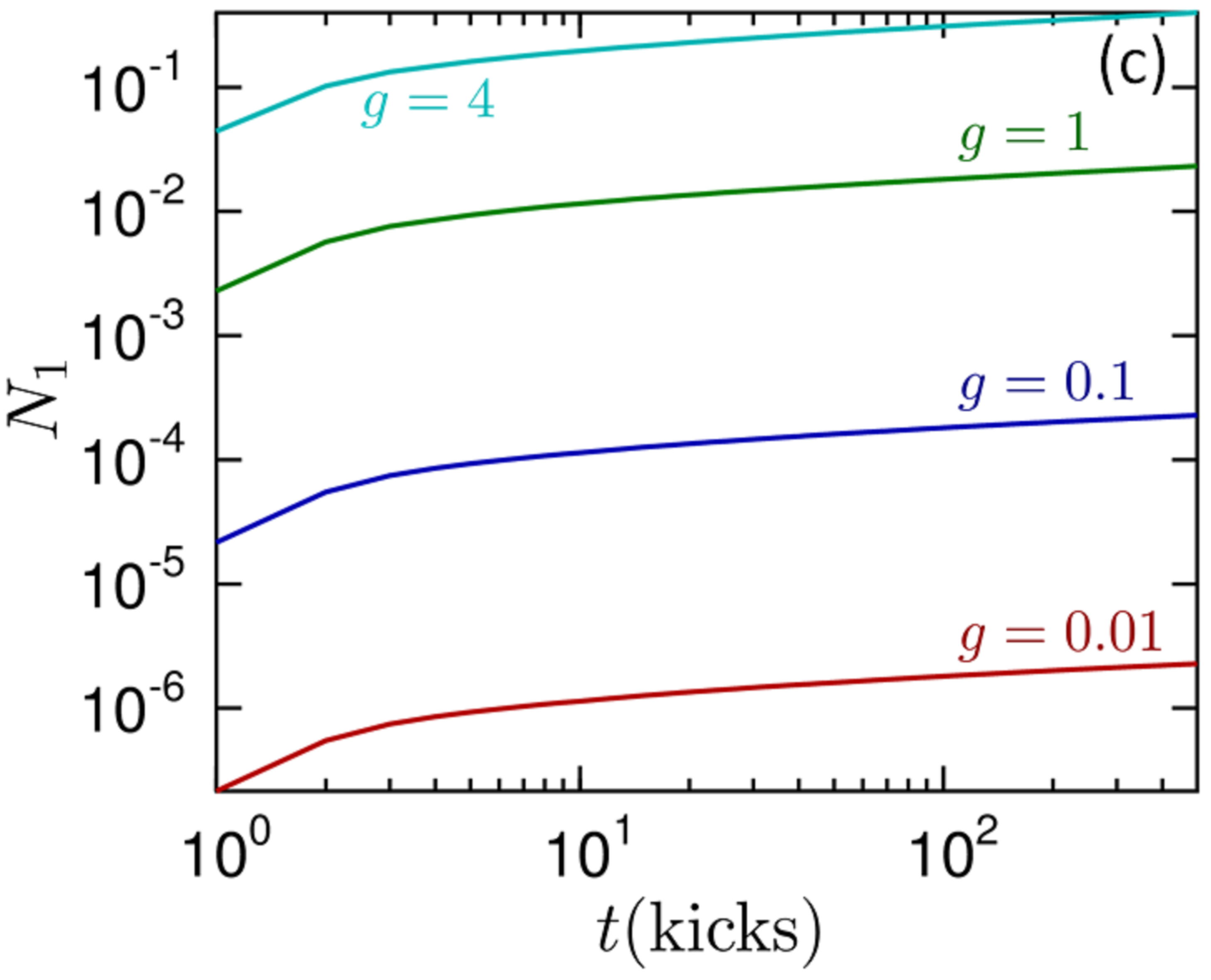}$\quad$\includegraphics[height=6cm]{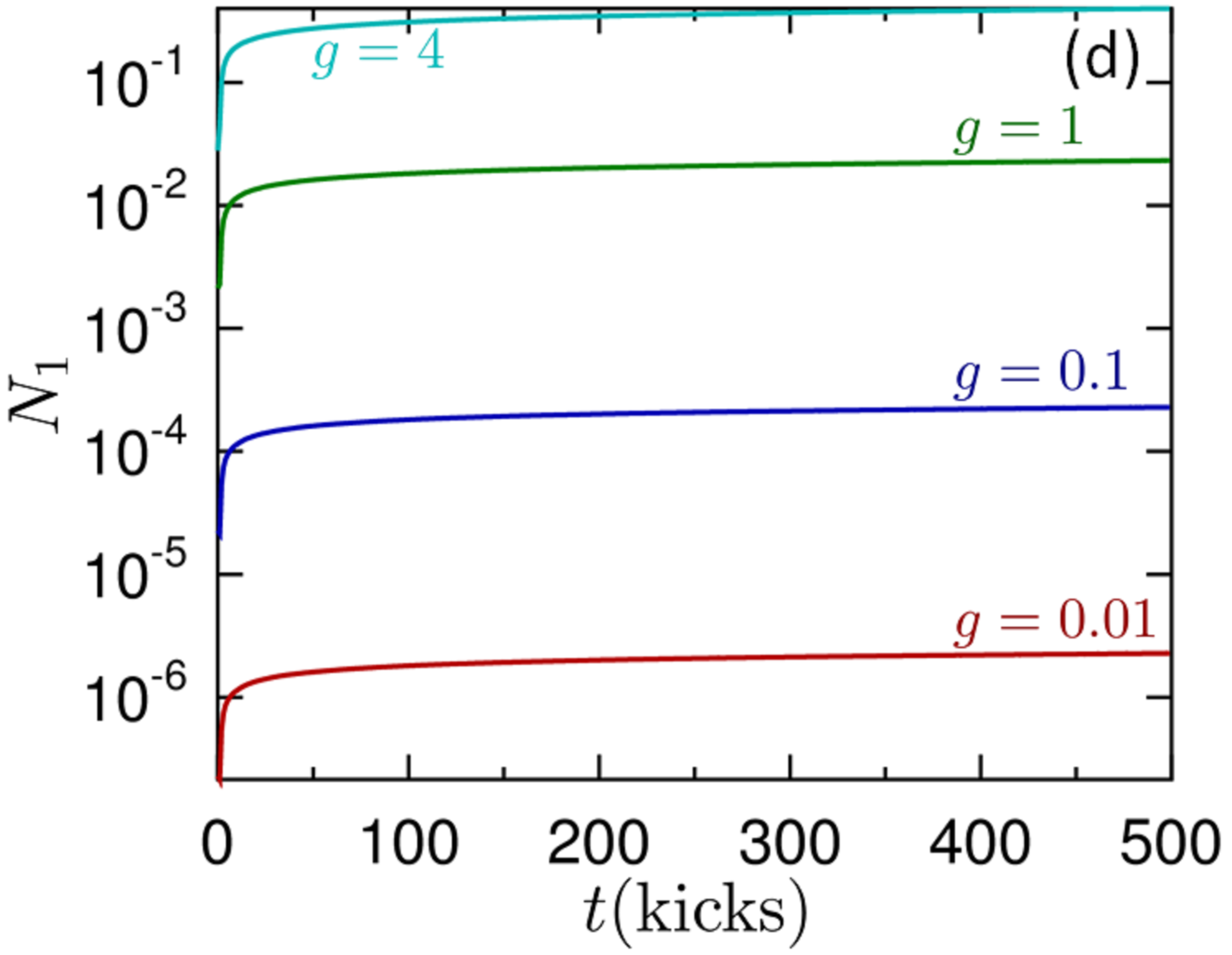}
\par\end{centering}
\caption{\label{fig:N1vst}Evolution of the excitation number $N_{1}$ of the $k=1$ Bogoliubov mode, in different regimes: quasi-insulator {[}(a) log-log and (b)
semilog scale{]}, $K=4,\,\varepsilon=0.1$ and metal {[}(c) log-log
and (d) semilog scale{]}, $K=9,\,\varepsilon=0.8$, for four values
of the interaction strength $g$ increasing from bottom to top: 0.01
(red, bottom curve), 0.1 (blue), 1 (green) and 4 (cyan, top curve).
Plot (a) shows a drastic growth of the excitation number for $g\ge1$,
indicating the onset of an instability, the dashed line in plot (a)
indicates the behavior in the absence of interactions (constant $N_{1}$).
Plot (b) (semilog scale) shows that for $g=4$ this growth is exponential.
For $g=1$ it is also roughly exponential, with fluctuations, but
for lower values of $g$ it is algebraic, approximately $\propto t$ on the considered
time scale of $10^{4}$ kicks. In plots (a) and (b), large values of $N_1$ (larger than the initial number of atoms in the condensate) are unphysical and signal the failure of the linearized Bogoliubov approach. In the metallic regime, no instability
is apparent up to 500 kicks.}
\end{figure*}

A kicked rotor is realized by submitting ultracold atoms to short
kicks of a standing wave at times separated by a constant interval
$T_{1}$. If such kicks have a constant amplitude, one obtains the
standard (periodic) kicked rotor which exhibits dynamical localization~\cite{Casati:LocDynFirst:LNP79,Moore:AtomOpticsRealizationQKR:PRL95},
i.e. localization in momentum space. If the amplitude of the kicks
is modulated with a quasiperiodic function
\mbox{$F(t)=1+\varepsilon\cos\left(\omega_{2}t+\varphi_{2}\right)\cos\left(\omega_{3}t+\varphi_{3}\right)$}
, where $\omega_{2}T_{1}$, $\omega_{3}T_{1}$ and $\kbar\equiv4\hbar k_{L}^{2}T_{1}/M$
(the reduced Planck constant) are incommensurable ($k_{L}$ is the
wave-vector of the standing wave and $M$ is the mass of the atoms),
the QPKR is obtained~\cite{Casati:IncommFreqsQKR:PRL89,Shepelyansky:3DRandomPot:JPI96}.
In the absence of particle-particle interactions, the QPKR Hamiltonian,
in conventional normalized units~\cite{Moore:AtomOpticsRealizationQKR:PRL95,Lemarie:AndersonLong:PRA09},
is: 
\begin{equation}
H(t)=\frac{p^{2}}{2}+K F(t)\cos x \sum_{n\in\mathbb{N}}\delta(t-n).\label{eq:Hqpkr}
\end{equation}
where $K$ is proportional to the average standing wave intensity.
In such units the time interval between kicks is $T_{1}=1$ and lengths
are measured in units of $(2k_{L})^{-1}$. Throughout this work we
take $\omega_{2}=2\pi\sqrt{5}$, $\omega_{3}=2\pi\sqrt{13}$ and $\kbar=2.89$
corresponding to typical experimental values~\cite{Chabe:Anderson:PRL08,Lemarie:CriticalStateAndersonTransition:PRL10,Lopez:ExperimentalTestOfUniversality:PRL12}.
In the absence of interactions the QPKR displays, for low values of
$K$ and $\epsilon$, dynamical localization at long times (i.e. $\left\langle p^{2}\right\rangle \sim\mathrm{constant}$);
for $K\gg1$, $\epsilon\approx1$ one observes a diffusive regime
$\left\langle p^{2}\right\rangle \sim t$, and in between there is
a critical region which displays a subdiffusive behavior $\left\langle p^{2}\right\rangle \sim t^{2/3}$~\cite{Lemarie:AndersonLong:PRA09}.
In the presence of weak interactions modeled as a mean-field nonlinear
potential, the critical and the diffusive regimes are not affected,
whereas the localized regime is replaced by a subdiffusive one $\langle p^{2}\rangle\sim t^{\alpha}$,
with $\alpha\sim0.4$~\cite{Cherroret:AndersonNonlinearInteractions:PRL14,Ermann:DestructionOfAndersonLocNonlinearity:JPAMG14}.
In the following, we consider low enough interaction strengths and
short enough times so that this change of behavior is not significant; we shall
thus use the term ``quasilocalized'' (or ``quasi-insulator'')
to characterize this phase.

We use in the present work a model that is slightly different from
the experimentally realized QPKR: We
use periodic boundary conditions over one spatial period of the optical potential. In such a case,
$p$ becomes a discrete variable $p=\kbar l$, with $l\in\mathbb{Z}$.
In this model there is no spatial dilution of the boson gas, and the average
nonlinear potential, which is proportional to the atom density, does
not vary with time. This is not the case in the usual experimental
realization of the QPKR, where the atom cloud diffuses with time in
\emph{momentum space} (so that, even in presence of dynamical localization,
it is still undergoing spatial dilution), causing a significant diminution
of the spatial density; once the system is diluted, the nonlinearity
does not play any important role. Our model is thus expected to catch
more clearly the physics in presence of the nonlinearity. Such a model
can be realized experimentally by using a tightly confined toroidal
trap~\cite{Ramanathan:SuperflowToroidalBEC:PRL11} formed by higher
order Laguerre-Gauss modes in which atoms are confined in the radial
direction but are free to rotate. The azimuthal dependence of such
modes can be used to create a sinusoidal intensity modulation \emph{along}
the torus, analogous to a standing wave~\cite{Ngcobo:DigitalLaserForOnDemandLaser:NCM13},
in the present case a superposition of $\mathrm{LG}_{01}$ and $\mathrm{LG}_{0-1}$
modes. Note that in such a geometry collective effects
can manifest themselves when interactions become strong~\cite{Dubessy:CriticalRotationAnnularBEC:PRA12},
but here we will be mainly interested in the weak interactions regime.

\begin{figure*}
\begin{centering}
\includegraphics[height=6cm]{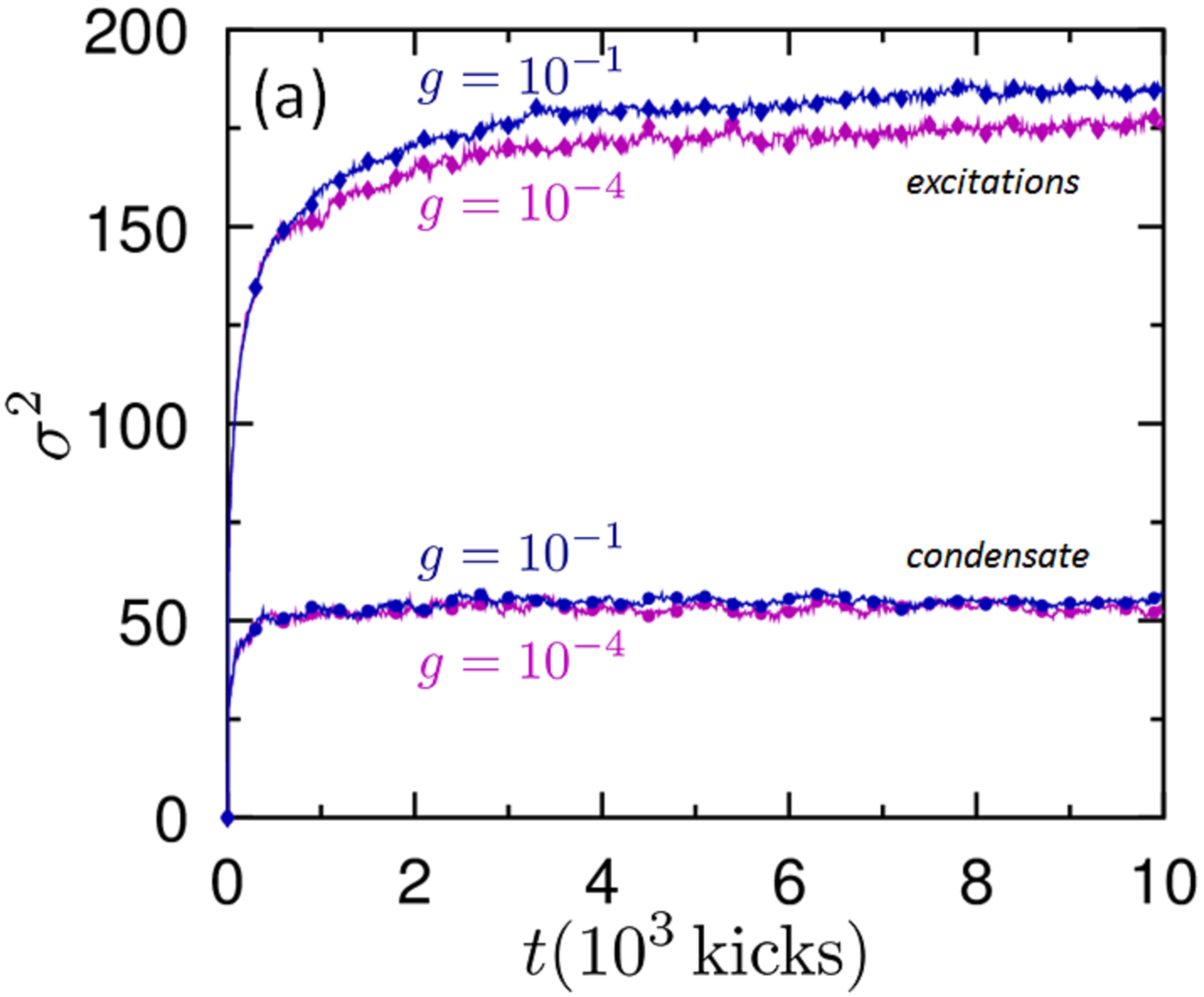}$\quad$\includegraphics[height=6cm]{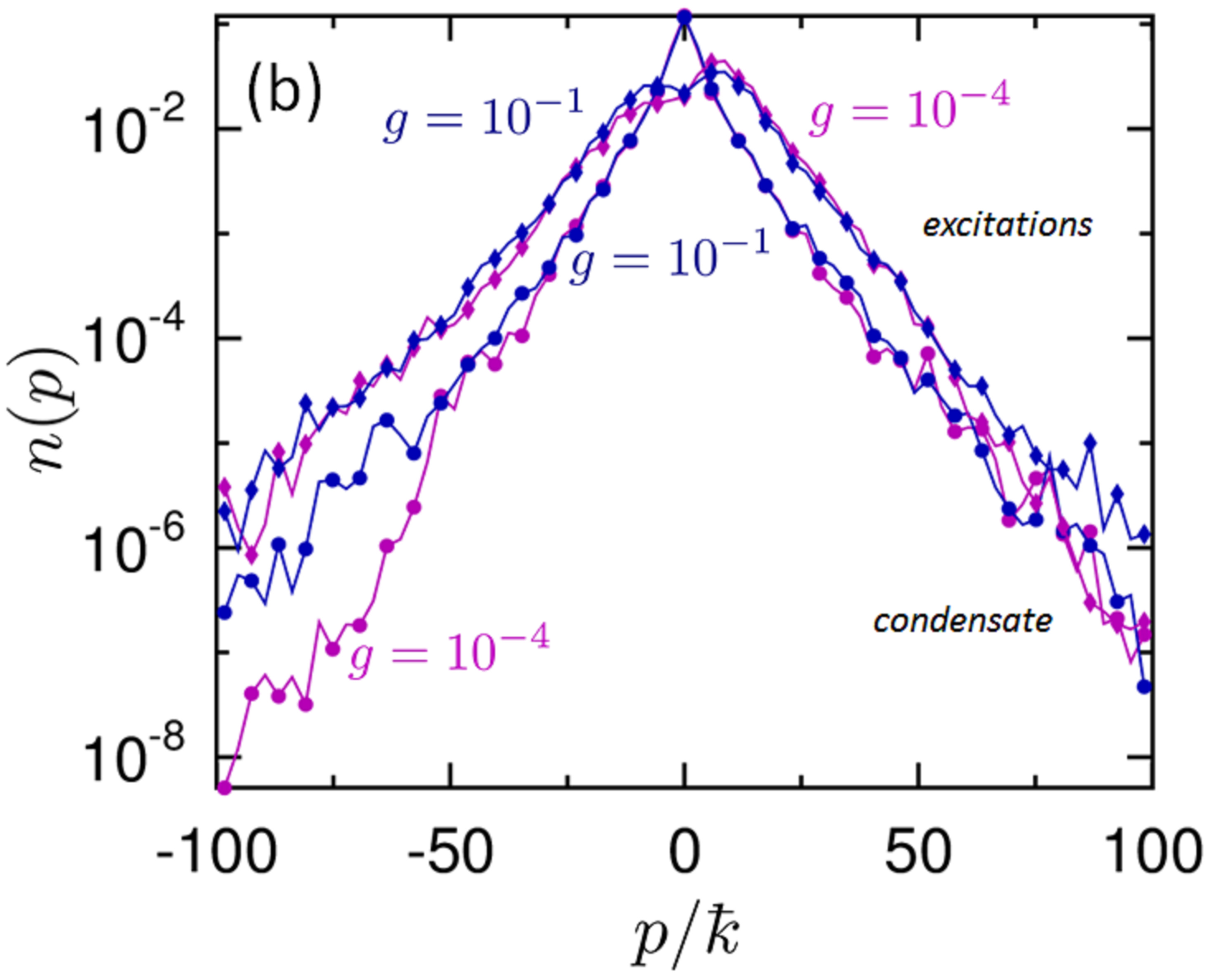}\\
\par\end{centering}
\centering{}\includegraphics[height=6cm]{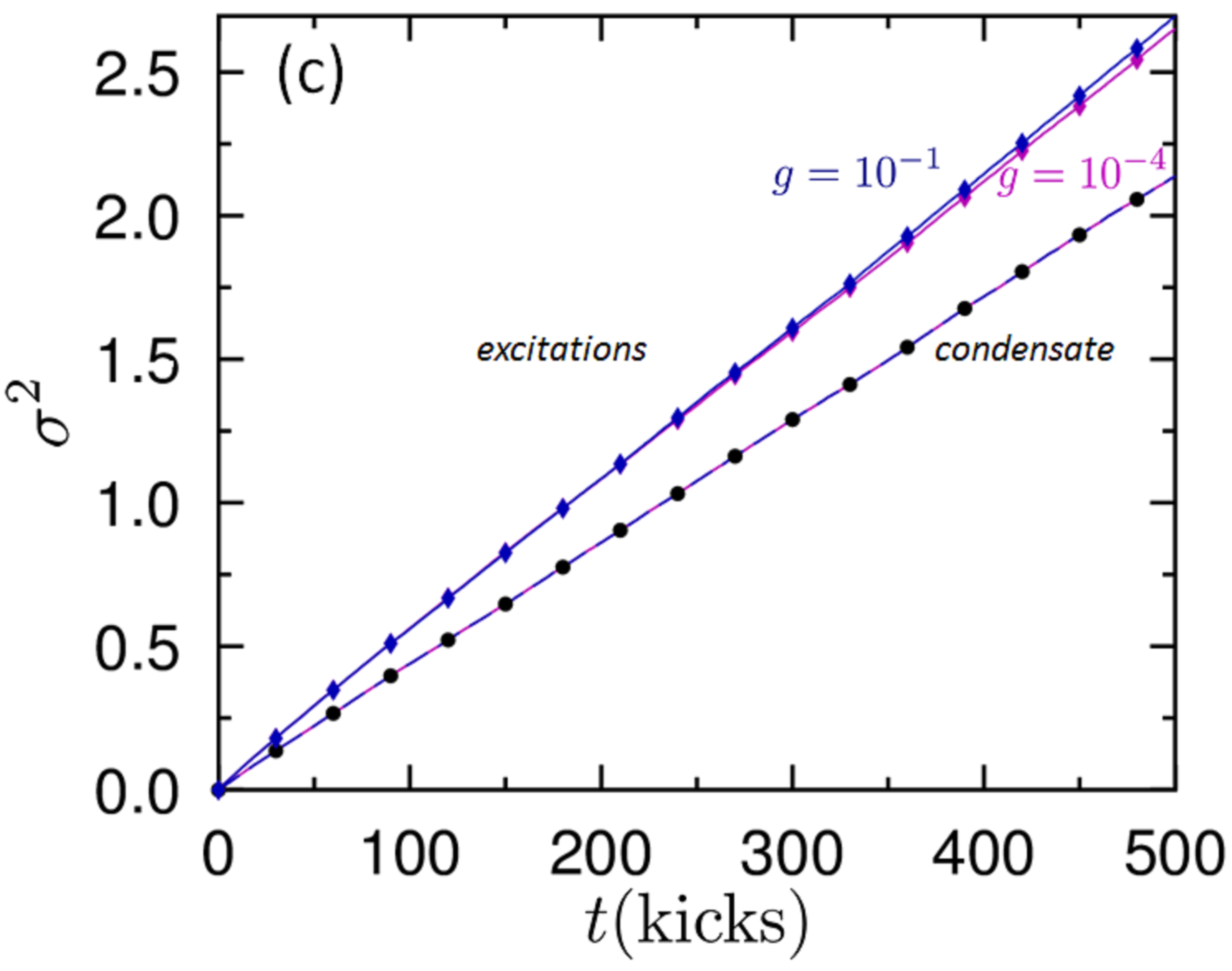}$\quad$\includegraphics[height=6cm]{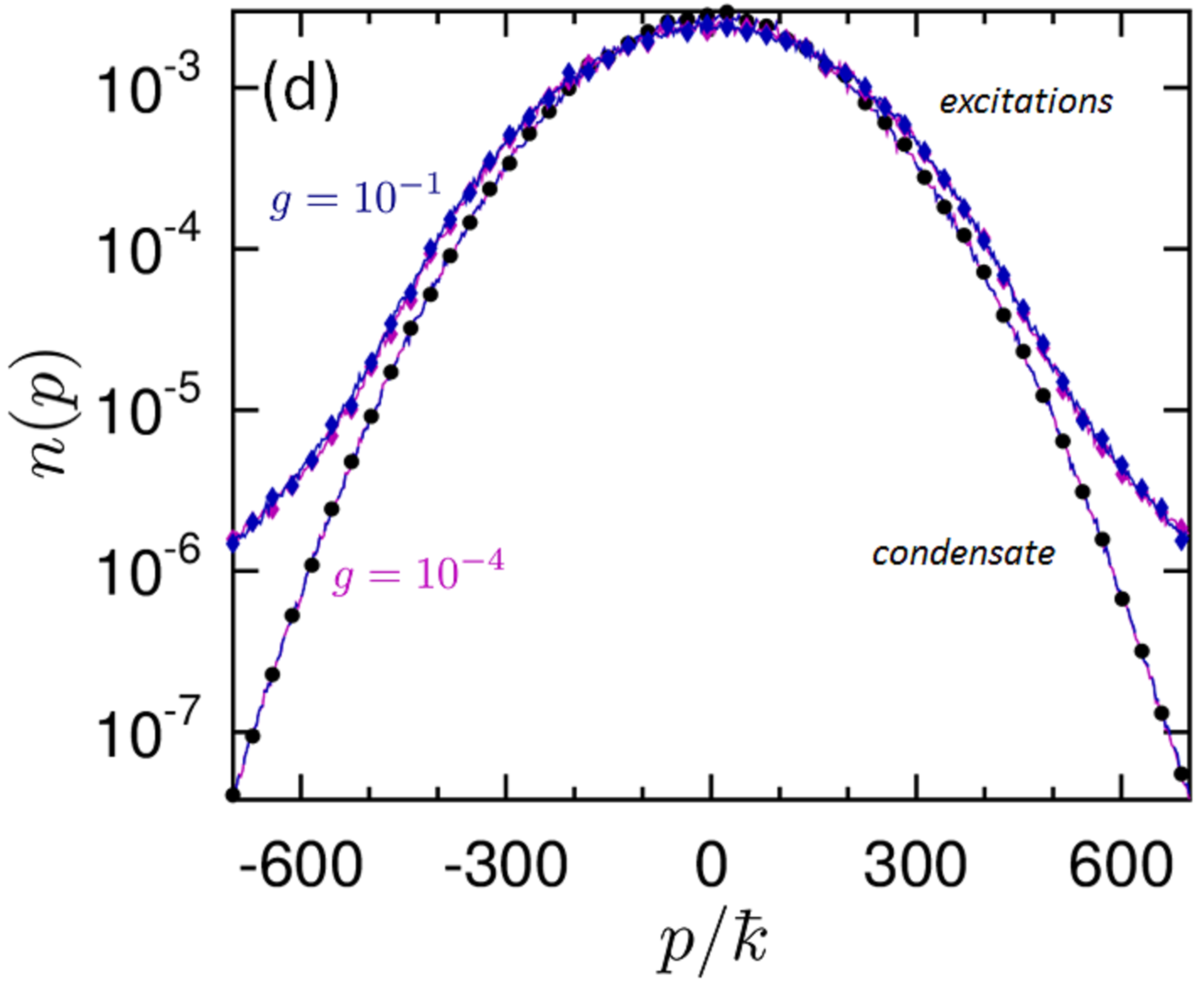}\caption{\label{fig:profiles}Dynamics of the condensate and of the Bogoliubov
excitations. The momentum variance $\sigma^{2}$ of the condensate
(circles) and of the excitations (diamonds) for $g=10^{-4}$ (magenta)
and $g=10^{-1}$ (blue) are shown in plot (a) for the quasilocalized
regime ($K=4,\varepsilon=0.1$, $t\le10^{4}$) and in plot (c) for
the diffusive regime ($K=9,\varepsilon=0.8$, $t\le500$). The momentum
distributions $n(p)$ of the condensate (circles) and of the excitations
(diamonds) in logarithmic scale are shown in (b) for the quasilocalized
regime (at $t=10^{4}$) and in (d) for the diffusive regime (at $t=500$). }
\end{figure*}

In this quasi-1D geometry, we take interactions into account via the
particle-number-conserving Bogoliubov formalism~\cite{Castin:LowTemperatureBECsInTimeDependent:PRA98},
at zero temperature. The gas of interacting bosons is separated into
two parts: i) The condensed fraction (the ``condensate'') and ii)
the non-condensed fraction (``excitations'' or ``quasiparticles'').
The condensate is governed by the Gross-Pitaevskii equation 
\begin{equation}
i\kbar\frac{\partial\phi(x,t)}{\partial t}=H(t)\phi(x,t)+g|\phi(x,t)|^{2}\phi(x,t)\label{eq:evolphi},
\end{equation}
where the condensate wave function $\phi$ is normalized to unity:
$\int_{0}^{2\pi}|\phi(x,t)|^{2}dx=1$ ($L=2\pi$ is the system length)
and the rescaled 1D interaction strength $g=2\kbar\omega_{\perp}aN$
is proportional to the $S$-wave scattering length $a$, the number
of atoms $N$ and the transverse trapping frequency $\omega_{\perp}$. 

The non-condensed part is described in the Bogoliubov formalism as
a set of \emph{independent} bosonic quasiparticles, with a two-component
state vector $(u_{k},v_{k})$ satisfying the normalization condition
\begin{equation}
\int_{0}^{2\pi}\left(|u_{k}(x,t)|^{2}-|v_{k}(x,t)|^{2}\right)dx=1,
\end{equation}
and evolving according to the equation: 
\begin{equation}
i\kbar\partial_{t}\left[\begin{array}{c}
u_{k}(x,t)\\
v_{k}(x,t)
\end{array}\right]=\mathcal{L}\left[\begin{array}{c}
u_{k}(x,t)\\
v_{k}(x,t)
\end{array}\right],\label{eq:evol_uk}
\end{equation}
where the operator $\mathcal{L}$ is a $2\times2$ matrix: 
\begin{eqnarray}
\mathcal{L} &\!=\!& \left[\begin{array}{cc}
Q(t) & 0\\
0 & Q(t)^{\dagger}
\end{array}\right]\mathcal{L_{\mathrm{GP}}}\left[\!\begin{array}{cc}
Q(t) & 0\\
0 & Q(t)^{\dagger}
\end{array}\!\right]\nonumber \\
\mathcal{L_{\mathrm{GP}}} &\!=\!& \left[\!\begin{array}{cc}
H+2g|\phi|^{2}\!-\!\mu(t) & g\phi^{2}\\
-g{\phi^*}^{2} & -H\!-\!2g|\phi|^{2}+\mu(t)
\end{array}\!\right],\nonumber
\end{eqnarray}
with $\mu(t)=\int_0^L \left(\phi^{*}H\phi+g|\phi|^{4}\right)dx$ the time-dependent
chemical potential. The presence of the projection operator $Q(t)=1-|\phi\rangle\langle\phi|$
ensures the conservation of the total number of particles~\cite{Castin:LowTemperatureBECsInTimeDependent:PRA98}.

The goal of this work is to study the (i) stability and the (ii) dynamical localization properties of a quasi-periodically kicked condensate.
The stability of the condensate can be assessed by monitoring the
number of non-condensed atoms (the quantum depletion) at zero temperature,
which is given by $\delta N=\sum_{k}N_{k}$, where
\begin{eqnarray}
N_{k} & = & \int_{0}^{2\pi}|v_{k}(x,t)|{}^{2}dx\nonumber
\end{eqnarray}
is the number of excitations in the mode $k$. To describe the localization properties of the system, we will expand the condensate wave function and the Bogoliubov mode in Fourier series 
\begin{eqnarray}
f(x,t) & = & \frac{1}{2\pi} \sum_{l\in\mathbb{Z}}e^{ilx}\tilde{f}(l,t)\\
\tilde{f}(l,t) & = & \int_{0}^{2\pi}e^{-ilx}f(x,t)dx,
\end{eqnarray}
where $f=\phi,u_{k},v_{k}$. The momentum distribution of the condensate and of Bogoliubov excitations (in a mode $k$) then read \mbox{$n_{c}(p=\kbar l)=|\tilde{\phi}(l)|^{^{2}}/(2\pi \kbar)$} and \mbox{$n_{b}(p=\kbar l)= \left|\tilde{v}_{k}(l)\right|^{2}/(2\pi N_{k}\kbar)$}, respectively.

Finally, for our numerical study we will take as initial conditions for the $(u_{k},v_{k})$ amplitudes the eigenstates of the operator $\mathcal{L}(t=0)$\footnote{The first kick is applied at $t=0^{+}$.}, which are plane waves of momentum $\kbar k$~\cite{Castin:LowTemperatureBECsInTimeDependent:PRA98}
\begin{equation}
\left[\begin{array}{c}
\tilde{u}_{k}(l,t=0)\\
\tilde{v}_{k}(l,t=0)
\end{array}\right]=\sqrt{\frac{\pi}{2}}\left[\begin{array}{c}
\zeta+1/\zeta\\
\zeta-1/\zeta
\end{array}\right]\delta_{k,l}\label{eq:uk0}
\end{equation}
with $k\in\mathbb{Z}^{*}$, and $\zeta$ given by: 
\begin{equation}
\zeta=\left[\frac{k^{2}}{k^{2}+2g/\pi\kbar^{2}}\right]^{1/4}.\label{eq:def_xi}
\end{equation}
In the example below, we will focus on the evolution of the $k=1$ Bogoliubov mode, which is initially the most populated,  see Eqs.~\eqref{eq:uk0},\eqref{eq:def_xi}~\footnote{
We have also checked numerically for parameters of the stable localized regime ($g=10^{-4}$, $K=4$, $\epsilon=0.1$) 
that the other modes $k\neq 1$ show similar behavior, and in particular that the $k=1$ mode remains the most populated one during the evolution.}.

\section{Stability of the condensate}

For the periodic kicked rotor, several studies showed the emergence
of an instability at large positive values of $g$ (repulsive interactions)
\cite{Zhang:TransitionToInstabilityKickedBEC:PRL04,Billam:CoherenceAndInstabilityInADrivenBEC:NJP12,Reslen:DynamicalInstabilityInKickedBEC:PRA08},
which manifests itself by an exponential increase of the number of
excitations. We shall now study this instability in the \textit{quasiperiodic}
kicked rotor for $g>0$. Equations~\eqref{eq:evolphi} and~\eqref{eq:evol_uk}
can be integrated simultaneously by a split-step method. Numerical
data are averaged over 500 random realizations of the phases $\varphi_{2},\varphi_{3}\in[0,2\pi)$. 
As the total number of particles is fixed, the number of condensed
particles is $N-\delta N$ and the non-condensed fraction $\delta N/N$.
As long as $\delta N$ is much smaller that the typical number of
atoms $\approx10^{5}$ used in a experiment, the kicked condensate will be considered to be stable.

Fig.~\ref{fig:N1vst} (a) and (b) display the onset of the instability
at a relatively large time (compared to experimentally accessible
time scales) of $10^{4}$ kicks in the quasi-insulator region $K=4, \varepsilon=0.1$.
For low $g$ values, the number of excitations $N_{1}$ increases
moderately with time, approximately $\propto t$. For higher values of $g$, an explosive
growth of $N_{1}$ indicates that the condensate has lost its stability
(and that the Bogoliubov approximation has lost its validity). However, for
$g\lesssim1$, even at the limit of an experiment duration $t=10^{3}$
the instability has not set in, which is confirmed by the small number
of excitations $N_{1}(g=1,t=10^{3})\approx9.5$. As shown in appendix~\ref{app},
the origin of the dynamical instability can be understood by studying
a simplified evolution operator illustrating the competition between
the kick and the interactions that can destroy the condensate. Plots
(c) and (d) show the evolution of the number of excitations in the
metal (diffusive) regime $K=9,\varepsilon=0.8$, up to $t=500$. In
this case, the condensate is less affected by the presence of interactions,
as the kinetic energy grows linearly with time and eventually dominates
the constant interaction energy $\simeq g/(2\pi)$ (see below). 

\section{The quasi-insulator-metal transition in the stable region}

We now focus on low interacting strengths $g\le0.1$ for which $N_{1}\ll N$,
meaning that the system is stable (up to the times considered here)
and the Bogoliubov formalism is valid. We will study the localizing
properties of the system to find whether the system follows the non-interacting
regimes (localized, diffusive) or if the localized phase is replaced
by a subdiffusive phase~\cite{Shepelyansky:KRNonlinear:PRL93,Cherroret:AndersonNonlinearInteractions:PRL14}.
We will also discuss how the critical properties of the transition
are changed by interactions.

For $g\le0.1$ and for short enough times, one expects the condensate
to display (quasi)localization if $K<K_{0}$ and diffusion if $K>K_{0}$,
$K_{0}$ being the critical point. Our numerical simulations show
that this is also the case for the excitations. Fig.~\ref{fig:profiles}(a) shows the second moment of the distribution
$\sigma_{i}^{2}=\langle p^{2}\rangle_{i}-\langle p\rangle_{i}^{2}$,
with $\langle p^{2}\rangle_{i}=\kbar^{3}\sum l^{2}n_{i}(p)$ and $\langle p\rangle_{i}=\kbar^{2}\sum_{l}ln_{i}(p)$, for
both the condensate ($i=c$, circles) and the excitations ($i=b$,
diamonds) in the quasilocalized regime. For the two values of the
interaction strength, $g=10^{-4}$ (magenta curve) and $g=10^{-1}$
(blue curve), the second moment of the condensate saturates to a constant
value $\sigma_{c}^{2}\approx50$, showing that the wave-packet is
quasilocalized. More interestingly, the curves with diamonds markers in
Fig.~\ref{fig:profiles}(a) show that variance of the momentum of
Bogoliubov excitations \emph{also tend to saturate}, with a larger
value $\sigma_{b}^{2}\approx180$: quasiparticles also (quasi-)localize.

The momentum distributions $n_{c}$ and $n_{b}$ in the quasilocalized
regime at $t=10^{4}$ for $g=10^{-4}$ are shown in Fig.~\ref{fig:profiles}(b)
{[}same graphical conventions as in Fig.~\ref{fig:profiles}(a){]}.
Both distributions remain essentially centered around the origin so that
their first moment $\langle p\rangle_{i}$
($i=c,b$) remains small, while they show an exponential behavior in the wings. 
For the condensate, assuming an exponential profile $n_{c}(p)\propto\exp(-|p|/\xi)$
[see Fig.~\ref{fig:profiles}(b)], the width $\xi$ of the momentum
distribution at $t=10^{4}$ is given by $\xi=\sigma_{c}/\sqrt{2}\approx5$,
which evolves very slowly up to $t=10^{4}$. Thus, for very weak interactions
and on the time range accessible to experiments, the condensate behaves
as a single particle and displays similar behaviors in the vicinity of the Anderson transition.
The Bogoliubov distribution presents a double peak near the center. 
This peculiar shape is probably due to the fact that the initial momentum distribution
of the mode $k=1$ {[}Eq.~\eqref{eq:uk0}{]} is centered at $p=\kbar$,
thus breaking the symmetry between positive and negative momenta.
The wings of the excitations momentum distribution and of the condensate
have approximately the same slope, confirming that excitations have
the same localization length in this case. The fact that the former
has a flatter top than the later explains why $\sigma_{b}$ is significantly
larger than $\sigma_{c}.$

Fig.~\ref{fig:profiles}(c) is the equivalent of Fig.~\ref{fig:profiles}(a) in the diffusive regime $K=9,\ \varepsilon=0.8$,
showing that $\sigma_{c}^{2}$ and $\sigma_{b}^{2}$ increase linearly
with time, with a diffusion coefficient that is similar for $g=10^{-4}$
and $g=10^{-1}$, $D_{c}=\sigma_{c}^{2}/2t\approx 20$ and $D_{b}=\sigma_{b}^{2}/2t\approx 25$.
Fig.~\ref{fig:profiles}(d) {[}equivalent to Fig.~\ref{fig:profiles}(b){]}
represents the corresponding momentum distributions at $t=500$. Both
have the typical Gaussian shape associated with a diffusion process.

These results show that when the condensate is stable and for
experimentally relevant times, the system is not affected by the presence
of (weak) interactions and that Bogoliubov excitations display the
same dynamics as particles. For larger interaction strengths,
one expects the presence of a subdiffusive phase instead of the localized
regime~\cite{Shepelyansky:KRNonlinear:PRL93,Gligoric:InteractionsDynLocQKR:EPL11,Rebuzzini:EffectsOfAtomicInteractionsQKR:PRA07,Cherroret:AndersonNonlinearInteractions:PRL14}.
However, for the range of parameters investigated in this work ($10^{-4}<g<4,K=4,\varepsilon=0.1$)
we found that the condensate is \emph{never stable and subdiffusive
at the same time}. For $g\gtrsim0.1$ interactions appear to be more
likely to destroy the condensate than to induce subdiffusion.

\begin{figure}
\begin{centering}
\includegraphics[height=6cm]{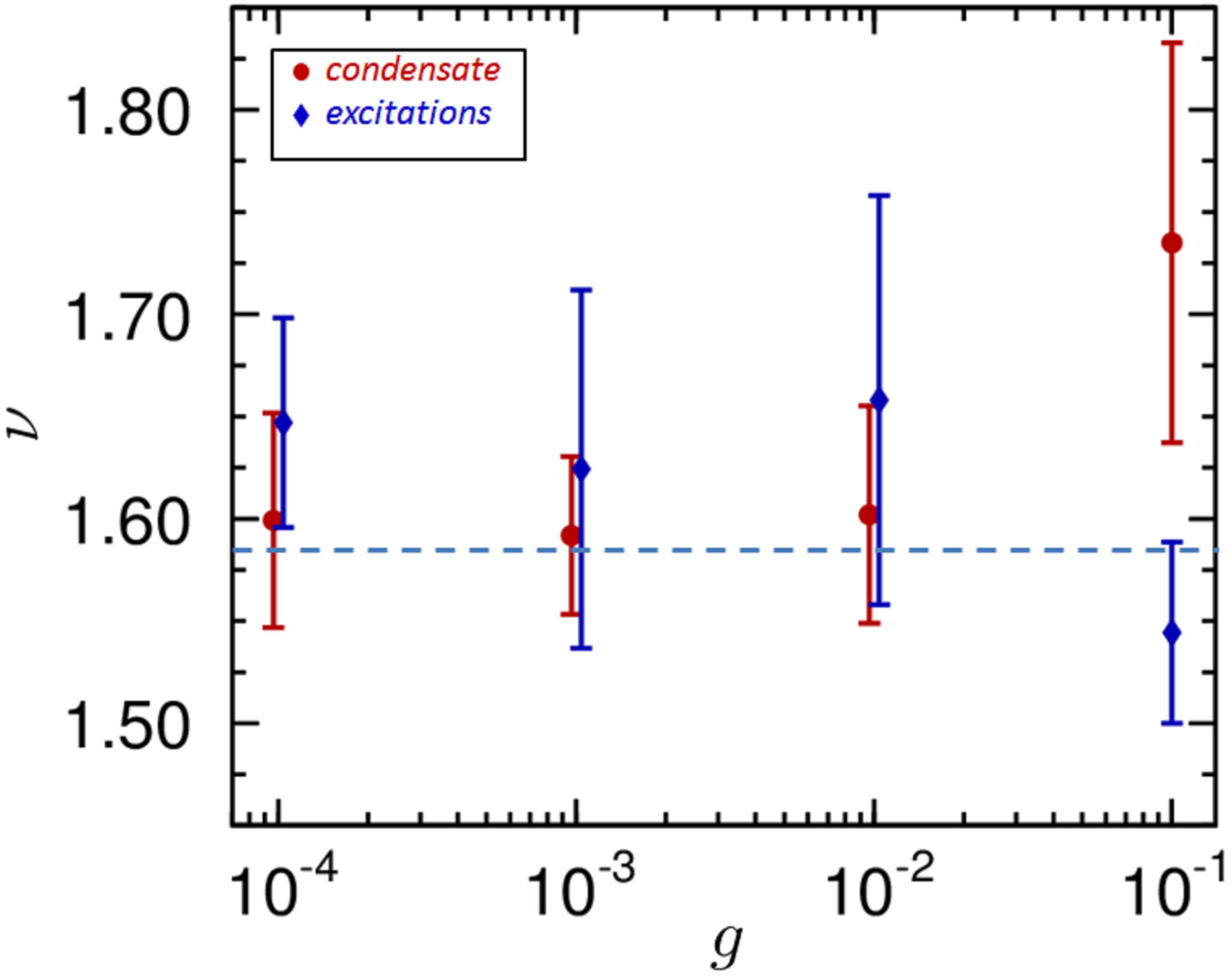}
\par\end{centering}
\caption{\label{fig:critical}Critical exponent $\nu$ vs interaction
strength $g$ for both the condensed fraction (red circles) and the
Bogoliubov excitations (blue diamonds). Error bars are calculated
via a standard bootstrap method~\cite{NumRec:07}. The points were
slightly shifted horizontally so that error bars do not superpose.
The blue dashed line indicates the critical exponent value $\nu\approx1.58$
in the absence of interactions.}
\end{figure}

The fact that Bogoliubov excitations behave like (non-interacting)
particles in the quasi-insulator and metal regimes in the stable region,
suggests that they display a quantum phase transition of the same
nature as the Anderson transition, which can be verified by determining
its (universal) critical exponent $\nu$. The universality of this
second-order phase-transition has been demonstrated experimentally
in the absence of interactions~\cite{Lopez:ExperimentalTestOfUniversality:PRL12},
giving an experimental value for the critical exponent $\nu=1.63\pm0.05$,
independent of microscopic parameters and consistent with the numerically
predicted value $1.58\pm0.02$~\cite{Lemarie:UnivAnderson:EPL09,Slevin:AndersonCriticalExp:NJP2014}.
We used a finite-time scaling method~\cite{Lemarie:AndersonLong:PRA09,Slevin:AndersonCriticalExp:NJP2014}
to extract a critical exponent $\nu$ from the dynamics of both the
condensate and the excitations. We chose the path in the parameter
plane $(K,\varepsilon)$ used in~\cite{Chabe:Anderson:PRL08}, $\varepsilon(K)=0.1+0.14(K-4)$.
Fig.~\ref{fig:critical} shows that, for small nonlinearities,
the critical exponent is the same for both components and compares
very well with the (non-interacting) experimental measurement, but
their values tend to become different for higher values of $g$. 
For $g\ge0.1$, the value of the critical exponent starts to deviate from
the universal value, as the system enters a new regime where the subdiffusive
character of the quasilocalized regime becomes important even for
short times~\cite{Cherroret:AndersonNonlinearInteractions:PRL14}. The
critical point is also the same for both the condensate and the excitations;
at $g=0$ its value is $K_{0}\approx6.38\pm0.05$ and changes only
slightly up to $g=0.1$, in accordance with the self-consistent theory
prediction of Ref.~\cite{Cherroret:AndersonNonlinearInteractions:PRL14}.
Hence, we can conclude that for low values of $g$, Bogoliubov excitations
undergo a second-order phase transition of the same nature as for non-interacting particles,
with the same critical exponent. 

\begin{figure*}
\centering{}\includegraphics[height=5cm]{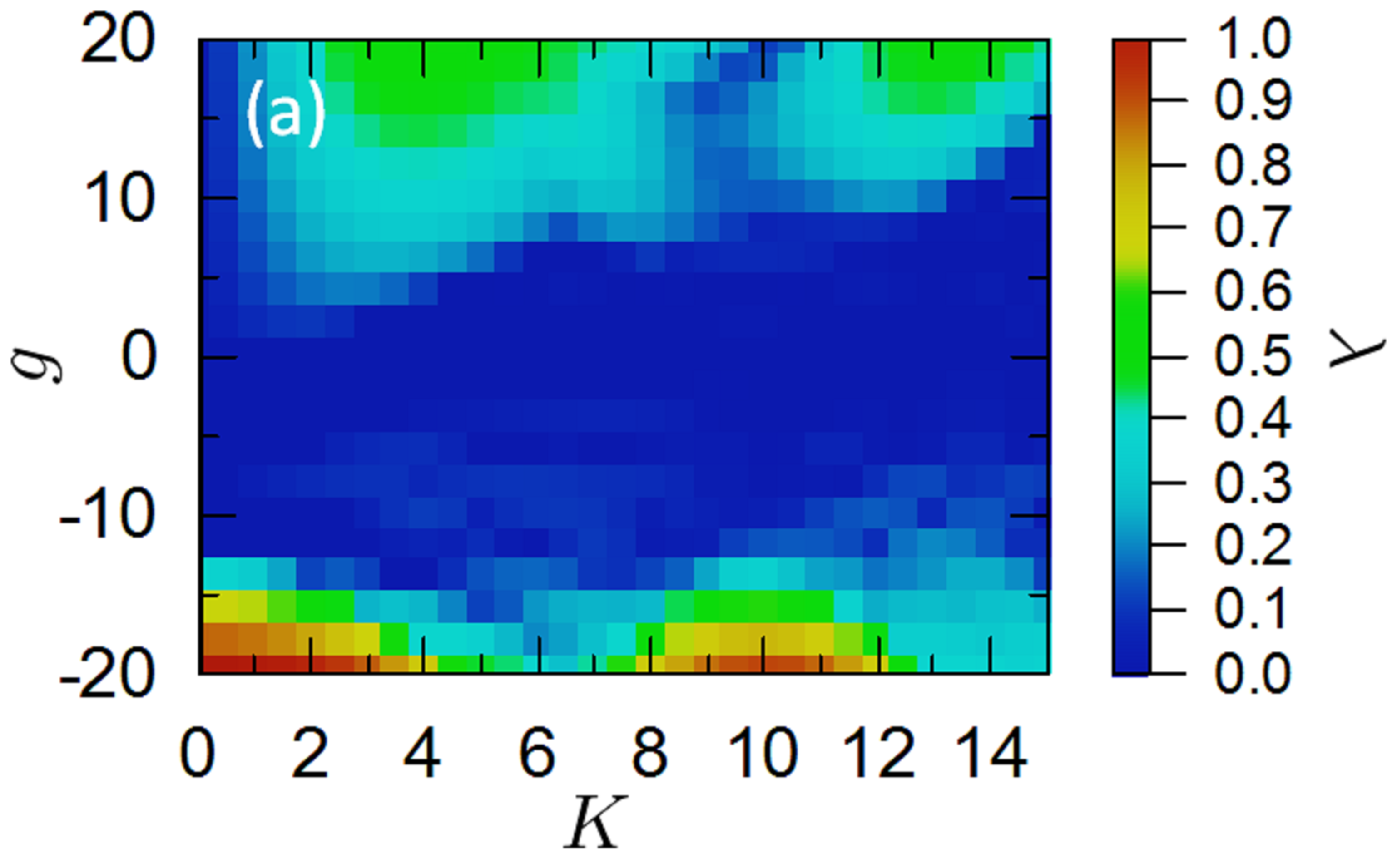}\includegraphics[height=5cm]{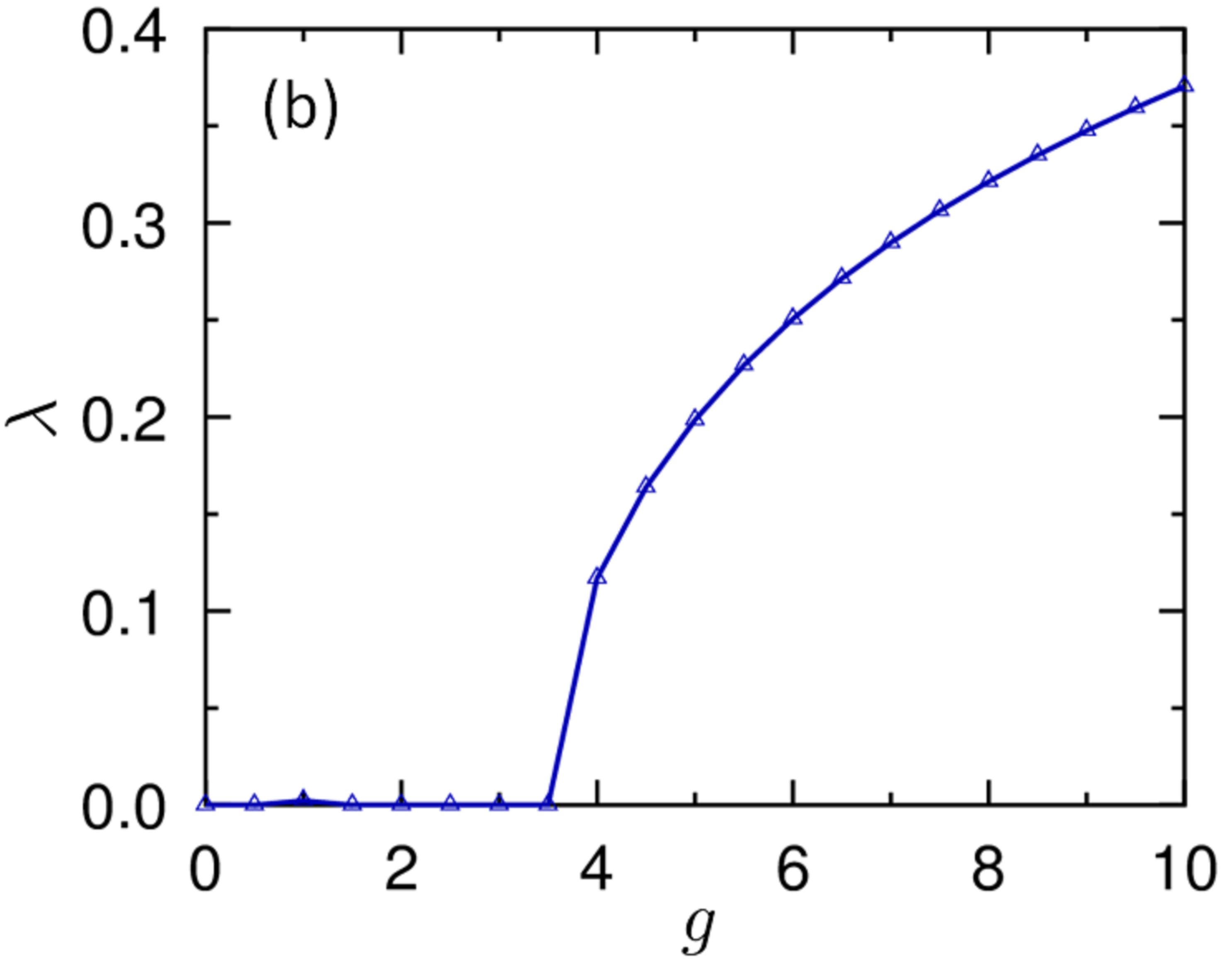}\caption{\label{fig:growthrate}(a) Largest growth rate $\lambda$ per kick
as a function of the kick amplitude $K$ and of the interaction strength
$g$, for $\kbar=2.89$. For $K=0$, the system is unstable for $g<-13$
as expected from Bogoliubov theory. For $K>0$, a dynamical instability
can also occur for $g>0$. For large negative $g$ the condensate is intrinsically instable, which explain the
non zero value of $\lambda$ even for $K=0$. Panel (b) shows the emergence of the instability
for $K=4$ and $g>0$. }
\end{figure*}

The previous analysis assumed that the system is prepared in its ground state [$\phi(x,t\!=\!0)=1/\sqrt{L}$], at zero temperature. 
The Bogoliubov modes are then initially populated only from quantum fluctuations. The very same equations of evolution~(\ref{eq:evol_uk}) -- with different initial conditions -- describe
the dynamics if the Bogoliubov modes are populated by some other process. 	
Experimentally, a specific Bogoliubov mode can be selectively excited using two laser waves
whose directions are chosen so that their wave-vector difference is equal
to the wave-vector $k$ of the desired mode~\cite{Steinhauer:ExcitationSpectrumBEC:PRL02}. Then the linear or exponential growth of the mode could be easily monitored experimentally.
The above study was restricted to the Bogoliubov mode $k=1$. Considering another mode $k\neq1$ is equivalent to a change of the
initial condition in the Bogoliubov equations. We checked numerically that other modes, display the same behavior, but
are much more affected by finite-time effects, as their initial momentum
distribution is more asymmetric {[}see Eq.~\eqref{eq:uk0}{]}.

\section{Conclusion}

In conclusion, in a quasiperiodic kicked rotor the condensate is stable
in the weakly interacting regime. For times longer than the
experimental time-scale (presently 1000 kicks), both the excitations
and the condensate display a behavior very close to the Anderson transition
of non-interacting particles, with the same critical exponent; the
universality of the phase transition is thus valid irrespective of
the nature of particles. However, our results also show that it might be difficult
to observe a subdiffusive phase at larger interacting strengths due
to the emergence of a dynamical instability. 
Thus, the fate of the transition in the presence
of strong interactions remains an open problem. For low positive values
of $g$ the transition can be experimentally observed, and, by increasing
interactions via a Feshbach resonance one can observe the onset of
nonlinear effects. This shows that the nonlinear regime can be ``approached
from below'' and that the transition can be observed within the stability
regime of the condensate. The present work paves the way for such
an experiment, which would represent an important step in our understanding
of interacting disordered systems presenting phase transitions.

\begin{acknowledgments}
The authors are grateful to Cord A. Mueller, R.~Chicireanu, J.-F.
Cl\'ement, P.~Szriftgiser, and N.~Cherroret for fruitful discussions.
This work is supported by Agence Nationale de la Recherche (Grants
LAKRIDI ANR-11-BS04-0003 and K-BEC No. ANR-13-BS04-0001-01), the Labex
CEMPI (Grant No. ANR-11-LABX-0007-01), Programme Investissements d'Avenir
under the program ANR-11-IDEX-0002-02, reference ANR-10-LABX-0037-NEXT,
and the Ministry of Higher Education and Research, Hauts de France
Council and European Regional Development Fund (ERDF) through the
Contrat de Projets Etat-Region (CPER Photonics for Society, P4S).
\end{acknowledgments}
\bigskip

\appendix

\section{Dynamical instability of the KR Bogoliubov modes\label{app}}

The goal of this appendix is to explain, in a simpler case, the features
of Fig.~\ref{fig:N1vst} by analyzing the properties of the operator
$\mathcal{L}$ in Eq.~\eqref{eq:evol_uk}. We use two important simplifications:
i) we neglect the influence of the modulation $\varepsilon$ considering
the standard kicked rotor single-particle Hamiltonian \eqref{eq:Hqpkr}
with $K(t)=K=\mathrm{constant}$ and ii) we make the assumption that the
condensate wavefunction is homogeneous, $\phi(x)=1/\sqrt{2\pi}$ so
that Eqs.~\eqref{eq:evol_uk} form a closed set of equations. The
projector $Q$ then becomes time-independent and the dynamical properties
of the system are governed by $\mathcal{L}_{\mathrm{GP}}$ only. In
order to study the influence of $g$ on the stability of the system,
we consider the evolution operator over one period: 
\begin{eqnarray}
U&=&\exp\left(-\frac{i}{\hbar}\left[\begin{array}{cc}
\frac{p^{2}}{2}+\frac{g}{2\pi} & \frac{g}{2\pi}\\
-\frac{g}{2\pi} & -\frac{p^{2}}{2}-\frac{g}{2\pi}
\end{array}\right]\right) \nonumber \\ 
&&\left[\begin{array}{cc}
U_{K} & 0\\
0 & U_{-K}
\end{array}\right].\label{eq:U}
\end{eqnarray}
In momentum space, the kick operator is: 
\begin{equation}
\langle p=\kbar l|U_{K}|p=\kbar m\rangle=(-i)^{(m-n)}J_{m-n}(K/\kbar).
\end{equation}
The eigenstates of $U$ contains all the dynamical properties of the
system: If an eigenstate of $U$ corresponds to a complex eigenvalue $\epsilon$
with $|\epsilon|>1$, the system will develop a dynamical instability
with a growth rate $\log|\epsilon|$. In Figure~\ref{fig:growthrate},
we represent the growth rate $\lambda=\log|\epsilon|$ associated
to the eigenstate with the largest eigenvalue $\epsilon$ as a function
of $K$ and $g$, for $\kbar=2.89$.

The limiting case $K=0$ provides a good test for our method as we
know from the standard Bogoliubov theory \cite{PethickSmith:BoseEinstein:08}
that the system is unstable for $g<-\kbar^{2}\pi/2\sim-13$. In presence
of the kicks $K>0$, the system can now develop a dynamical instability
for repulsive interactions $g>0$. For $K=4$, we find that the system
is indeed unstable at large $g$ which is in qualitative agreement with the full
numerical result of Figure \ref{fig:N1vst}(a). The critical value $g_c\approx 3.5$ for the instability is however overestimated, due to the strong assumptions
i) and ii). We also find that the system tends to be more stable at
large kick amplitudes, which is compatible with the result shown in
Fig.~\ref{fig:N1vst}(b). Finally in the non-interacting limit obtained
for $g=0$, the system is stable as the eigenvalues of the evolution
operator, Eq.~\eqref{eq:U} coincides with those of the standard
non-interacting kicked rotor. In particular, the results obtained above are compatible
with those of Ref.~\cite{Liu:TransitionToInstabilityKickedBECRing:PRA06}. 


%

\end{document}